\documentclass{elsart}
\usepackage{graphicx}

\begin{document}

\begin{frontmatter}

\title{Lateral Distribution of the Radio Signal in Extensive Air Showers Measured with LOPES}

\author[fzk]{W.D.~Apel},
\author[uni]{J.C. Arteaga\thanksref{r0}},
\author[ipe]{T.~Asch},
\author[fzk]{A.F.~Badea},
\author[nij]{L.~B\"ahren},
\author[fzk]{K.~Bekk},
\author[tor]{M.~Bertaina},
\author[mpi]{P.L.~Biermann},
\author[fzk,uni]{J.~Bl\"umer},
\author[fzk]{H.~Bozdog},
\author[buc]{I.M.~Brancus},
\author[sie]{M.~Br\"uggemann},
\author[sie]{P.~Buchholz},
\author[nij]{S.~Buitink},
\author[tor,toi]{E.~Cantoni},
\author[tor]{A.~Chiavassa},
\author[uni]{F.~Cossavella},
\author[fzk]{K.~Daumiller},
\author[uni]{V.~de Souza\thanksref{r1}},
\author[tor]{F.~Di~Pierro},
\author[fzk]{P.~Doll},
\author[fzk]{R.~Engel},
\author[nij,ast]{H.~Falcke},
\author[fzk]{M. Finger}, 
\author[wup]{D.~Fuhrmann},
\author[ipe]{H.~Gemmeke},
\author[toi]{P.L.~Ghia}, 
\author[wup]{R.~Glasstetter},
\author[sie]{C.~Grupen},
\author[fzk]{A.~Haungs},
\author[fzk]{D.~Heck},
\author[nij]{J.R.~H\"orandel},
\author[nij]{A.~Horneffer},
\author[fzk]{T.~Huege},
\author[fzk]{P.G.~Isar},
\author[wup]{K.-H.~Kampert},
\author[uni]{D. Kang}, 
\author[sie]{D.~Kickelbick},
\author[ipe]{O.~Kr\"omer},
\author[nij]{J.~Kuijpers},
\author[nij]{S.~Lafebre},
\author[pol]{P.~{\L}uczak},
\author[uni]{M.~Ludwig},
\author[fzk]{H.J.~Mathes},
\author[fzk]{H.J.~Mayer},
\author[uni]{M.~Melissas},
\author[buc]{B.~Mitrica},
\author[toi]{C.~Morello},
\author[tor]{G.~Navarra},
\author[fzk]{S.~Nehls\thanksref{corr}},
\author[nij]{A.~Nigl},
\author[fzk]{J.~Oehlschl\"ager},
\author[sie]{S.~Over},
\author[uni]{N.~Palmieri},
\author[buc]{M.~Petcu},
\author[fzk]{T.~Pierog},
\author[wup]{J.~Rautenberg},
\author[fzk]{H.~Rebel},
\author[fzk]{M.~Roth},
\author[buc]{A.~Saftoiu},
\author[fzk]{H.~Schieler},
\author[ipe]{A.~Schmidt},
\author[fzk]{F.~Schr\"oder},
\author[ubu]{O.~Sima},
\author[nij]{K.~Singh\thanksref{r2}},
\author[buc]{G.~Toma},
\author[toi]{G.C.~Trinchero},
\author[fzk]{H.~Ulrich},
\author[fzk]{A.~Weindl},
\author[fzk]{J.~Wochele},
\author[fzk]{M.~Wommer},
\author[pol]{J.~Zabierowski},
\author[mpi]{J.A.~Zensus} 

\address[fzk]{Institut f\"ur Kernphysik, Forschungszentrum Karlsruhe, Germany}
\address[uni]{Institut f\"ur Experimentelle Kernphysik, Universit\"at Karlsruhe, Germany}
\address[ipe]{IPE, Forschungszentrum Karlsruhe, Germany}
\address[nij]{Department of Astrophysics, Radboud University Nijmegen, The Netherlands}
\address[tor]{Dipartimento di Fisica Generale dell' Universita Torino, Italy}
\address[mpi]{Max-Planck-Institut f\"ur Radioastronomie Bonn, Germany}
\address[buc]{National Institute of Physics and Nuclear Engineering, Bucharest, Romania}
\address[sie]{Fachbereich Physik, Universit\"at Siegen, Germany}
\address[toi]{Istituto di Fisica dello Spazio Interplanetario, INAF Torino, Italy}
\address[ast]{ASTRON, Dwingeloo, The Netherlands}
\address[wup]{Fachbereich Physik, Universit\"at Wuppertal, Germany}
\address[pol]{Soltan Institute for Nuclear Studies, Lodz, Poland}
\address[ubu]{Department of Physics, University of Bucharest, Bucharest, Romania}

\thanks[corr]{corresponding author, {\it E-mail address:} nehls@ik.fzk.de}
\thanks[r0]{now at: Universidad Michoacana, Morelia, Mexico}
\thanks[r1]{now at: Universidade S$\tilde{a}$o Paulo, Instituto de Fisica de S\~ao Carlos, Brasil}
\thanks[r2]{now at: KVI, University of Groningen, The Netherlands}

\begin{abstract}
The antenna array LOPES is set up at the location of the KASCADE-Grande extensive air shower experiment in Karlsruhe, Germany and aims to measure and investigate radio pulses from Extensive Air Showers. 
The coincident measurements allow us to reconstruct the electric field strength at observation level in dependence of general EAS parameters. 
In the present work, the lateral distribution of the radio signal in air showers is studied in detail. 
It is found that the lateral distributions of the electric field strengths in 
individual EAS can be described by an exponential function. 
For about 20\% of the events a flattening towards the shower axis is observed, 
preferentially for showers with large inclination angle.
The estimated scale parameters $R_0$, describing the slope of the lateral profiles range between   
$100$ and $200\,$m. No evidence for a direct correlation of $R_0$ with shower parameters like 
azimuth angle, geomagnetic angle, or primary energy can be found.  
This indicates that the lateral profile is an intrinsic property of the 
radio emission during the shower development which makes the radio detection technique 
suitable for large scale applications.
\end{abstract}

\end{frontmatter}

\section{Introduction}

The traditional method to study extensive air showers (EAS), which 
are generated by high-energy cosmic rays entering the Earth's atmosphere, 
is to measure the secondary particles with sufficiently large particle 
detector arrays. In general, these measurements provide only 
immediate information on the status of the air shower cascade 
at the particular observation level. This hampers the determination 
of the properties of the primary inducing the EAS as compared to 
methods like the observation of Cherenkov and fluorescence light, 
which also provide information on the longitudinal EAS 
development, thus providing a more reliable access to the 
information of interest~\cite{rpp}. 

In order to reduce the statistical and systematic uncertainties of 
the detection and reconstruction of EAS, especially with respect 
to the detection of cosmic particles of highest energies, measuring 
the radio emission during the shower development is being discusses 
as a new detection technique. 
The radio waves can be recorded day and night, and provide a 
bolometric measure of the electromagnetic shower component. 
Due to technical restrictions in past times, the radio emission accompanying 
cosmic ray air showers was a somewhat neglected feature in the past. 
However, the study of this EAS component has experienced a 
revival by recent activities, in particular by the LOPES 
project~\cite{Falcke05,Horn06}. LOPES as pathfinder for large scale radio 
detection for the LOFAR project~\cite{lofar} and the Pierre Auger Observatory~\cite{auger} 
investigates the correlations of radio data with shower parameters reconstructed by 
the extensive air shower experiment KASCADE-Grande~\cite{navarra}. 
Hence, LOPES, which is designed as a digital radio interferometer using 
large bandwidth and fast data processing, profits from the 
reconstructed air shower observables of KASCADE-Grande.

The main goal of the investigations in the frame of  
LOPES is the understanding of the shower radio emission 
in the primary energy range of $10^{16}\,$eV to $10^{18}\,$eV. 
Of particular interest is the investigation of the correlation 
of the measured field strength with main shower parameters. These are 
the orientation of the shower axis (azimuth angle, zenith angle, and the 
geomagnetic angle, i.e.~the angle between shower axis and geomagnetic field), 
the position of the observer relative to the shower axis, and the energy 
and mass (electron and muon number) of the primary particle. 
Another goal of LOPES (LOPES$^{\rm STAR}$) is the optimization of the hardware 
(antenna design and electronics) for a large scale application of the 
detection technique including a self-trigger mechanism for a stand-alone 
radio operation~\cite{Asch07}. 

In the present study we investigate in detail the lateral profile of the 
radio signal as measured by LOPES. Due to a 
precise amplitude calibration of each individual antenna and the event 
information from KASCADE-Grande, this is possible on an event-by-event basis 
with high accuracy. 
Such investigations are of great interest as the lateral shape defines 
the optimum grid size for a radio antenna array in a stand-alone mode.
Of particular interest is the scale parameter which describes the amount 
of the signal decrease with distance from the shower axis 
and the dependence of that parameter on characteristics of the primary particle. 
In addition, knowing the lateral extension in detail will contribute to the understanding of the emission mechanism of the radio signal as simulations have shown~\cite{Huege08}, that the lateral shape can be
related to important physical quantities such as the primary energy or the mass of the primary.  

\section{The LOPES Experiment}

\subsection{Experimental Setup}
\begin{figure}[ht]
 \centering
  \includegraphics[width=0.6\textwidth]{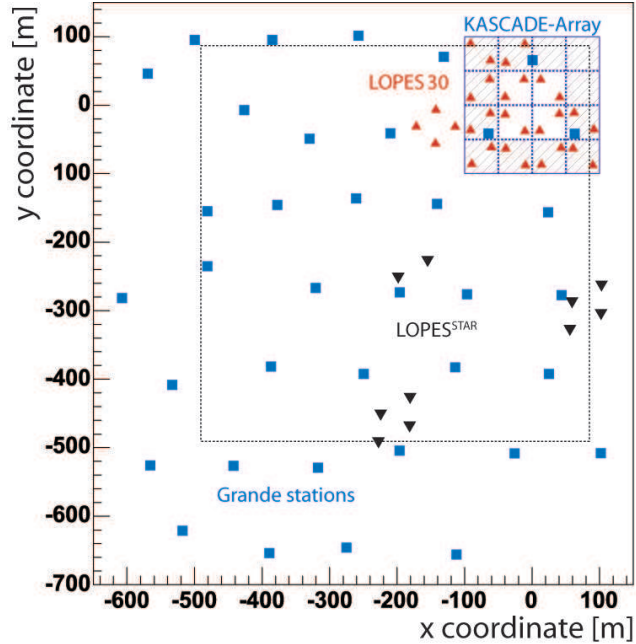}
\caption{Sketch of the KASCADE-Grande -- LOPES 
experiments: The squared 16 clusters (12 with muon counters) of the 
KASCADE field array, the distribution of the 37 stations of the Grande
array, the location of the 30 LOPES radio antennas, and the positions 
of the 10 newly developed LOPES$^{\rm STAR}$ antennas. The dotted line 
shows the area used for the present analysis.}
\label{FigLay}
\end{figure}

LOPES has been set-up as prototype station of the LOFAR project to verify 
the detection of radio emission from air showers. 
The basic idea is to use an array of relatively simple, quasi omni-directional 
dipole antennas. The  signals are digitized and sent to a central computer in 
which the registered waves from the individual antennas are superimposed 
(software interferometer). 
With LOPES it is possible to buffer the received data stream for a 
certain period of time. After a detection of a transient 
phenomenon like an air shower, a beam in the desired direction 
can be formed in retrospect.
To demonstrate the capability to measure air showers with 
such antennas, LOPES is situated
at the air shower experiment KASCADE-Grande. 
KASCADE-Grande is an extension of the multi-detector setup 
KASCADE~\cite{kascade}
(KArlsruhe Shower Core and Array DEtector) built in Germany, 
measuring the charged particles of air showers in the primary 
energy range of $10^{14}\,$eV to $10^{18}\,$eV with high precision due to 
the detection of the electromagnetic and the muonic shower component 
separately with independent detector systems. 
Hence, on the one hand LOPES profits from the reconstructed air 
shower observables of KASCADE-Grande, but on the other hand, since 
radio emission arises from different phases of the EAS development, 
LOPES is intended to provide complementary 
information to the particle detector arrays of KASCADE-Grande.

The antenna configuration of LOPES has changed several times in order to study
different aspects of the radio emission. With the extension of the antenna field
from 10 to 30 east-west orientated antennas in 2005 (LOPES30), the baseline 
and the low noise amplifier (LNA) performance at the antennas improved. 
In the configuration used in the analysis described here, 
LOPES operated 30 short dipole radio antennas. 
The LOPES antennas, positioned within or close to the original 
KASCADE array (fig.~\ref{FigLay}), operate in the frequency range of 
$40-80\,$MHz and are aligned in east-west direction, i.e. they are mainly 
sensitive to the linear east-west polarized component of the radiation.
This layout was in particular chosen to provide the possibility for a
detailed investigation of the lateral extension of the radio 
signal as it has a maximum baseline of approximately~$260\,$m.
The read-out window for each antenna is $0.8\,$ms wide centered around the trigger 
received from the KASCADE array; the sampling rate is $80\,$MHz. 
The logical condition to trigger the LOPES data readout is a high multiplicity of
fired particle detector stations of the KASCADE array. 
This corresponds to primary energies above $\approx 10^{16}\,$eV; 
which are detected with a rate of $\approx 2$ per minute. 

\subsection{Data Calibration}

Each single LOPES radio antenna has an absolute amplitude calibration at its 
location inside the KASCADE-Array (end-to-end calibration), performed using a 
commercial reference antenna~\cite{Nehls08} of known electric 
field strength at a certain distance. 
The power to be received from the source in calibration mode is compared with 
the power recorded in the LOPES electronics. 
The calibration procedure leads to frequency dependent amplification
factors representing the complete system behavior (antenna, cables, electronics)
in the environment of the KASCADE-Grande experiment. 
These correction factors are applied to the measured signal strengths resulting 
in electric field strengths which are used for further analysis. 
The systematic uncertainty of the calibration method (${\rm sys_{calib}}=20.5$~\%) 
is estimated from repeated measurement campaigns for individual
antennas under all kinds of conditions. This uncertainty also
includes, e.g., environmental effects, like those caused by different
weather conditions over the two years of calibration
campaigns. In particular, the antenna gain simulation contributes with an error
of $\approx 10$\% to the given total uncertainty~\cite{Nehls08}. 

The delay calibration of the system consists basically of two parts. 
First, the delay for each electronic channel is periodically verified 
with laboratory measurements of individual components and solar burst 
measurements of the complete signal chain.  
This delay was found to be almost stable over the operation time of LOPES. 
Second, a fine tuned correction of these delay times by a phase calibration is 
performed in order to catch short-time fluctuations in the trigger signal 
distribution and the data read-out.
The delay calibration is always performed in reference to one antenna
and the absolute time is connected to the data using the time stamps of 
KASCADE-Grande events. 
LOPES is operating three independent sub-stations with front-end
electronics and requires the trigger from KASCADE-Grande in each station.  
By that, a jitter due to long transmitting intra-station
cables is introduced. On basis of a known mono-frequent reference 
source with constant phase relation these disturbances are corrected. 
For the data sample used in this analysis, a TV-transmitter visible in the measured 
frequency range is used for the phase calibration~\cite{Horn06}.

\subsection{Pulse Height Calculations}

The LOPES data processing includes several steps. 
First, the relative instrumental delays are corrected using the 
known TV-transmitter.
Next, the gain corrections, digital filtering, and corrections 
of the trigger delays based on the known shower direction 
(from KASCADE-Grande) are applied and noisy antennas are flagged. 
These steps include the application of the gain and phase calibration 
correction factors and also a correction for the azimuth and zenith
dependence of the antenna gain. 

The first step of the beam forming procedure~\cite{Horn06} 
is the application of 
geometrical shifts of the data by the time difference 
of the pulse coming from the given direction to reach 
the position of the corresponding antenna.  
These shifts take into account the curvature of the electro-magnetic 
shower front (as free parameter to be fitted) and are done by multiplying a phase gradient 
in the frequency domain before transforming the data back to the time domain.
To form the beam, the data from each 
pair of antennas are multiplied time-bin by time-bin, the resulting 
values are averaged, and then the square root is taken while 
preserving the sign: $
cc(t) = \pm\sqrt{\left|\frac{1}{N_{\rm P}}
                 \sum_{i=1}^{N-1}\sum_{j>i}^{N}
                 s_i(t)s_j(t)\right|
                }$.
From the number of unique pairs of antennas $N_{\rm P}=(N-1)N/2$
the normalization is determined. The coherence of the
single antenna signals systematically influences the height and sign 
of this quantity, henceforth called the cc-beam~\cite{Horn06}. 
\begin{figure}[ht]
  \centering \includegraphics[width=0.48\textwidth]{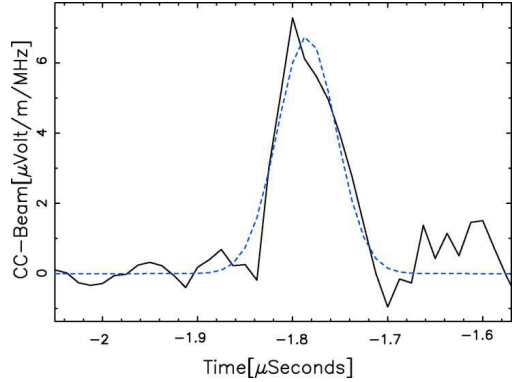}
  \caption{\label{ccbeam15-fit}
    The calculated cc-beam for an EAS is fit with a Gaussian function
    after block-averaging with a window of 37.5~ns.  The height
    [$\mu$V/m/MHz] and the center in time [$\mu$s] of the fit are used
    as parameters for the radio analysis.}
\end{figure}
\begin{figure}[ht]
  \centering \includegraphics[width=0.96\textwidth]{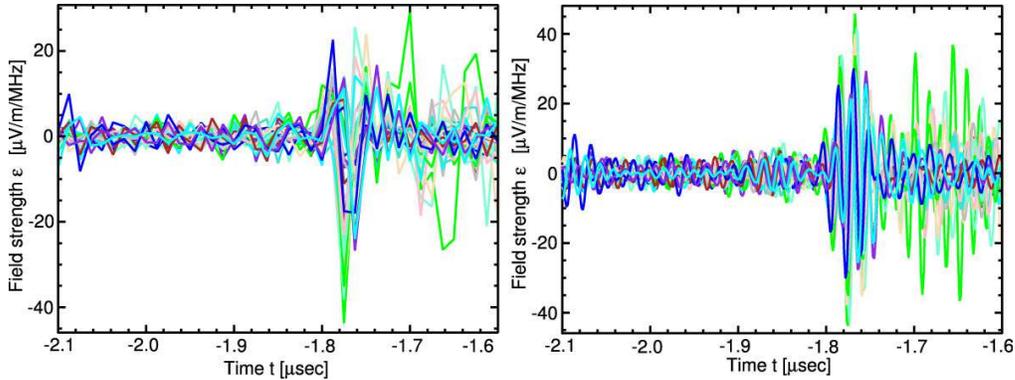}
  \caption{\label{upsample}
    Left: The sampled data with 12.5~ns spacing of the digitization
    with 80~MHz, in the second Nyquist domain. Right: The up-sampled
    signal shape between 40~MHz and 80~MHz.}
\end{figure}
In other words, the cc-beam is sensitive to and reflects the
coherence of the radio pulse. 
The quantification of the radio parameters is done by fitting the 
smoothed (over three time-samples block averaged $\equiv 1/\Delta \nu$) 
cc-beam pulse, as shown in figure~\ref{ccbeam15-fit}.
In first approximation the shape of the resulting pulse can be fitted 
by a Gaussian giving a robust value for the peak strength, 
which is defined as the height of this Gaussian. 
The error of the fit results gives also a first estimate of the 
uncertainty of this parameter. 
The finally obtained value $\epsilon_{\rm cc}$, which is the 
measured amplitude divided by the effective bandwidth, has
usually been used for LOPES analyses. 

While the cc-beam is an averaged property of the measured shower from all 
antennas, the present studies require the investigation of the field strength 
at individual antennas.
The sampling of the data is done in the second Nyquist domain but a
reconstruction of the original 40--80~MHz signal
shape is needed to estimate the field strength per antenna.
Therefore, an up-sampling of the data on a single antenna basis is performed 
(by the zero-padding method\footnote{Zero-padding is a method to extend a 
time series or a spectrum. Zeros are added in one domain and
after Fourier transformation an interpolated series is obtained in the 
other domain.
For the radio data the zero-padding is applied in the frequency domain and gives
a band limited interpolation in the time domain. The up-sampling of a data set
with N samples leads to a new data set with M samples.  
The up-sampling rate is given by $M/N = 2^n$, with $n = 0, 1, 2,...\,$.})
resulting in a 
band-limited interpolation in the time domain~\cite{Nehls08a} 
to reconstruct the original signal form between the sampled data 
points with 12.5~ns spacing. 
The method can be applied, because all needed information
after sampling in the second Nyquist domain is contained in the
stored data~\cite{Kroem08}. 
An example how the method reconstructs the original signal shape is
shown in figure~\ref{upsample}, where an up-sampling with $n=3$ is used. 

After applying the
zero-padding to the LOPES data, the radio signals can be used to
reconstruct the electric field strength in each individual antenna.
The Gaussian fit to the cc-beam defines 
the center of a time window with 45 nanoseconds width\footnote{For technical reasons 
the cc-beam calculation could not be used on the up-sampled data at 
the time of this analysis. This is not affecting the results of the 
present study, as the cc-beam is used only for the selection of the 
events and the definition of the time window.}. 
Within this time window, in the up-sampled data of each individual antenna 
the maximum absolute field strength is searched. This maximum value is used as 
the measured electric field strength $\epsilon$ per antenna and event. 
The window width is chosen to exclude the RFI appearing soon after the radio 
pulse signal. The RFI is caused by high currents in the particle 
detector cables during the penetration of the EAS.
The systematic uncertainty of $\epsilon$ is estimated to 19\%,
stemming mainly from the calibration procedure. 

As a further source of uncertainty, the contribution of the background
to the signal is taken into account. 
To evaluate this noise, the mean of the absolute electric field strength from 
a defined time window (520~ns width ending 300~ns before the observed 
radio pulse) is calculated, where a typical value is in the order of 15-20\% of the peak value of $\left|\epsilon\right|$ (see fig.~\ref{upsample}). 
As one does not know if the noise is constructively or destructively 
interfering with the signal, the value of the noise level is not subtracted 
from $\epsilon$, but added to the systematic uncertainty of the field strength.

\subsection{Event Selection}

The data set taken in the period from 16 November 2005 until 8 December 2006 is 
used for the analysis presented here, where 
for the whole period the 30 LOPES antennas have been
orientated in east-west direction.  
The data taking was interrupted in Summer 2006. 
During a thunderstorm on 27$^{th}$ of July an over-voltage in the trigger 
distribution system occurred, breaking the trigger electronics. The whole antenna field
was back in data acquisition in September 2006.  
Shorter malfunctions of individual antennas are handled by flagging them and 
not using them in the analysis.
In the relevant period, KASCADE generated 966,000 triggers that were sent to the LOPES
DAQ-system. Roughly 10\% of them could not be recorded due to the $1.5\,$s long 
dead-time for the read-out of the radio data from the memory boards.  
\begin{figure}[ht]
  \centering
  \includegraphics[width=0.48\textwidth]{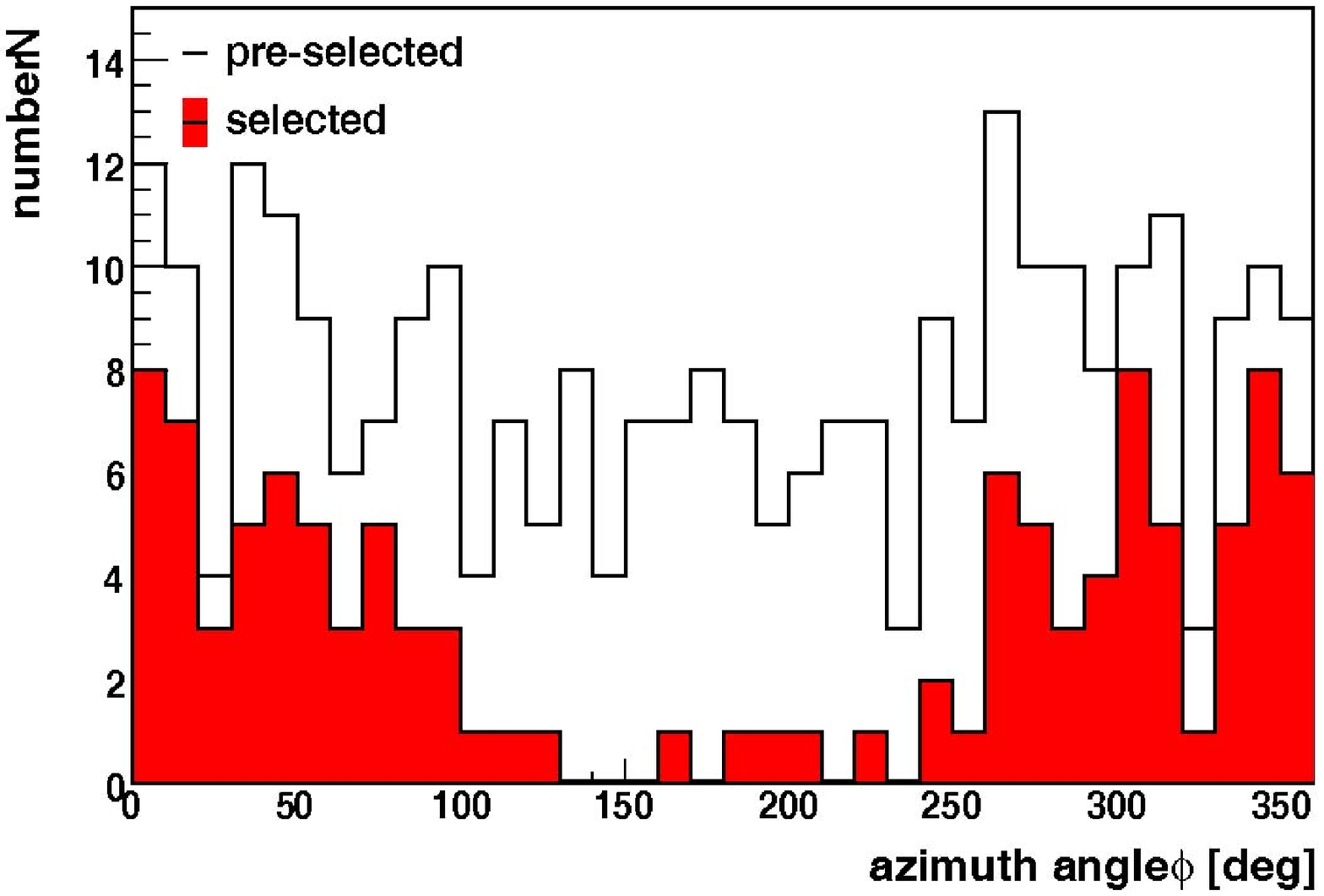}
  \includegraphics[width=0.48\textwidth]{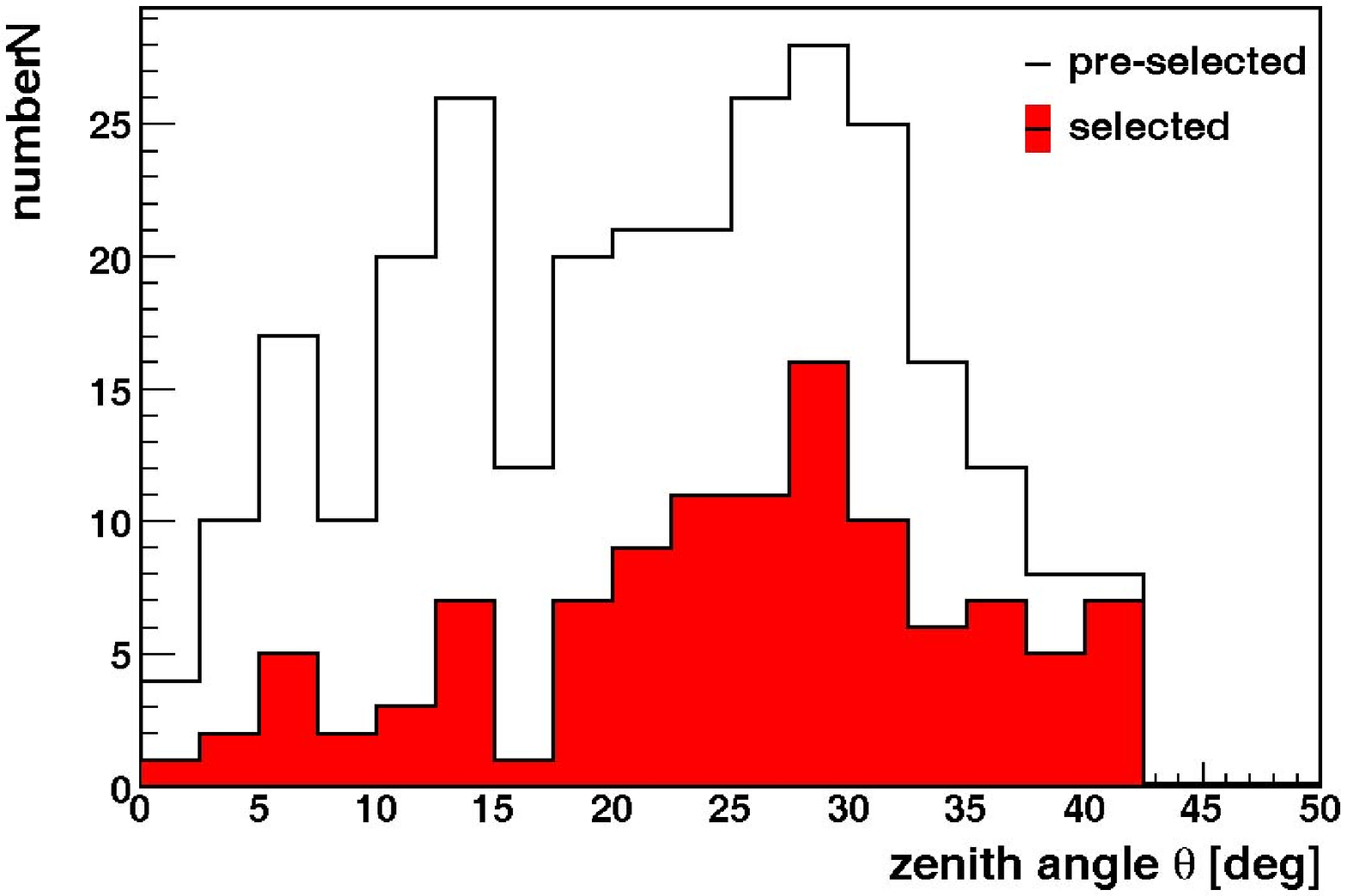}
  \includegraphics[width=0.48\textwidth]{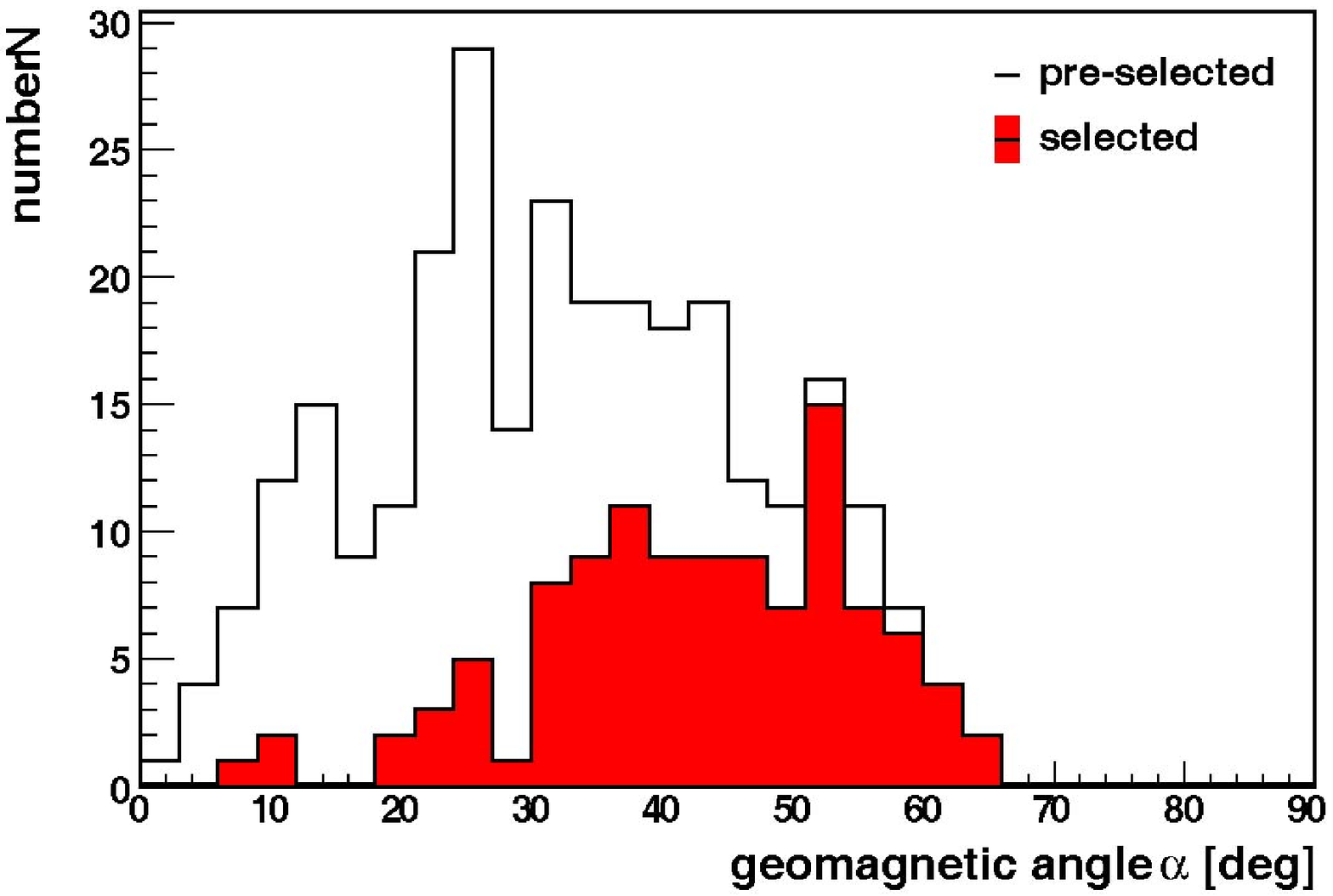}
  \includegraphics[width=0.48\textwidth]{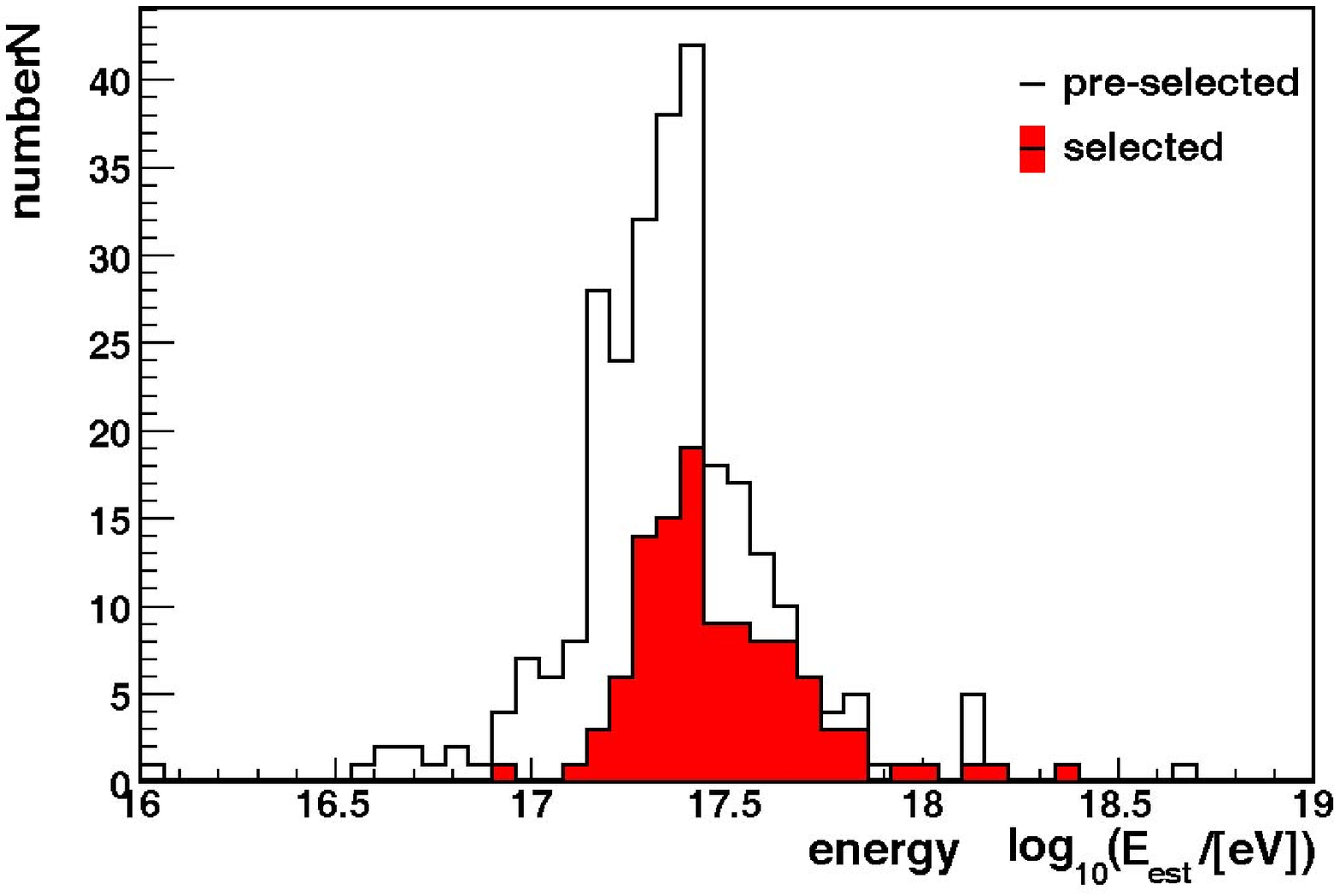}
  \caption{\label{selection}
    Distribution of the azimuth angle ($0 \equiv\,$North, $90 \equiv\,$East), 
    zenith angle, the geomagnetic angle, and
    the estimated primary energy $E_{\rm est}$ for all events selected by 
    KASCADE-Grande parameters (pre-selected) and for the events with clear 
    radio signals further considered in the analysis.}
\end{figure}
The general event reconstruction was performed for the remaining 860,000 events. 
Only a small fraction of this large amount of data shows radio signals
significantly above the background due to the high RFI at the 
KASCADE-Grande site~\cite{Badea06}. 
Only the highest energy EAS can give promising
radio signals well above the noise needed for the investigation of lateral 
distributions in individual events. 
To select those, several cuts are applied:
Firstly, the position of the shower axis has to be inside KASCADE-Grande, 
but not too far away from the antennas. 
Therefore, a square area of about 0.3~km$^2$ 
is used (see fig.~\ref{FigLay}).
Secondly, the shower parameters are properly reconstructed from the 
KASCADE-Grande data, and the shower energy exceeds $3-5\cdot10^{16}$~eV 
based on cuts at the total electron and muon number of the shower.

The number of events selected by these cuts is $N=296$. Twelve events
have been found above a zenith angle of $44^\circ$, and are not used due to a 
generally more uncertain KASCADE-Grande reconstruction for inclined events. 
The radio data of the selected events are first analyzed with 
the procedures described above in order to find events with significant 
radio signals above the background. 
The criteria for such events are a successful cc-beam fit 
(which is fulfilled for nearly all selected high-energy events except a few 
arriving directly from South) 
and a radio signal higher than the noise per individual, 
not flagged, up-sampled antenna in a $45\,$ns window around the 
center of the cc-beam pulse. 
These requirements result in a total number of $110$ events.

The geometrical layout of the 30 antenna array and the applied selection area define 
the mean distance of the shower axis to the antennas ($70-250\,$m; larger distances are 
suppressed due to the requirement on a clearly detected radio signal in all antennas) 
and the maximum baseline in individual events. 
A coverage of $100-300\,$m of the lateral distribution is obtained depending on the 
location of the shower core (see fig.~\ref{FigLay}).

In figure~\ref{selection}, the 110 radio events are compared to the original $284$ events 
selected by KASCADE-Grande observables, only in their general air shower parameters. 
The azimuth angles of all pre-selected events are almost uniformly distributed over the
whole angular range. An azimuth angle of $180^\circ$ corresponds to showers arriving from 
the South. For this direction the relative angle to the direction of the geomagnetic field 
(geomagnetic angle) is smaller than for showers coming from the 
North\footnote{The orientation of the magnetic 
field at the LOPES location is given by $\theta_b=25^\circ$ and $\phi_b=180^\circ$.
Showers from the South are almost parallel to the magnetic field and the resulting
Lorentz force is very small.}. 
If the magnetic field lines are nearly parallel to the shower axis,
the radio emission is expected to be generally much weaker and its polarization more 
directed towards the north-south axis~\cite{Falcke05,Huege07}. 
Therefore, as we measure only the east-west polarized component of the radio signal, 
no showers arriving from South fulfill the strict requirements to enter 
the further analysis. 
As an aside, this fact is a strong indication for a geomagnetic dependence of the radio emission. 
The zenith angle distribution has a mean of $22^\circ$, which is in agreement with the 
distribution of zenith angles of all KASCADE-Grande reconstructed events. 
Above $35^\circ$ zenith angle the fraction of radio events increases, 
as inclined showers have larger geomagnetic angles and therefore a larger
signal-to-noise ratio compared to nearly 
vertical showers, being closer to the orientation of the Earth's magnetic field.
In addition, inclined showers show a larger footprint on ground and the noise from the 
particle detectors is smaller.
At a fixed location the distribution of the geomagnetic angle 
incorporates the azimuth and zenith angle distributions, as both angles have to be 
used to calculate the geomagnetic angle $\alpha$.

The primary energy $E_{\rm est}$ is roughly
estimated from the KASCADE reconstructed muon and electron numbers with an accuracy 
of $\log(E_{\rm est}$/GeV$)\approx\pm0.15$. The arrival direction resolution of 
KASCADE-Grande is better than one degree for these high-energy events. 
The obtained distribution of the
estimated primary energy is also shown in figure~\ref{selection}, where 
the majority of the selected EAS have energies above $10^{17.2}\,$eV.

\section{Lateral Distributions of the Radio Signal in EAS}

\subsection{Reconstruction}
\begin{figure}[ht]
\begin{center}
  \includegraphics[width=0.48\textwidth]{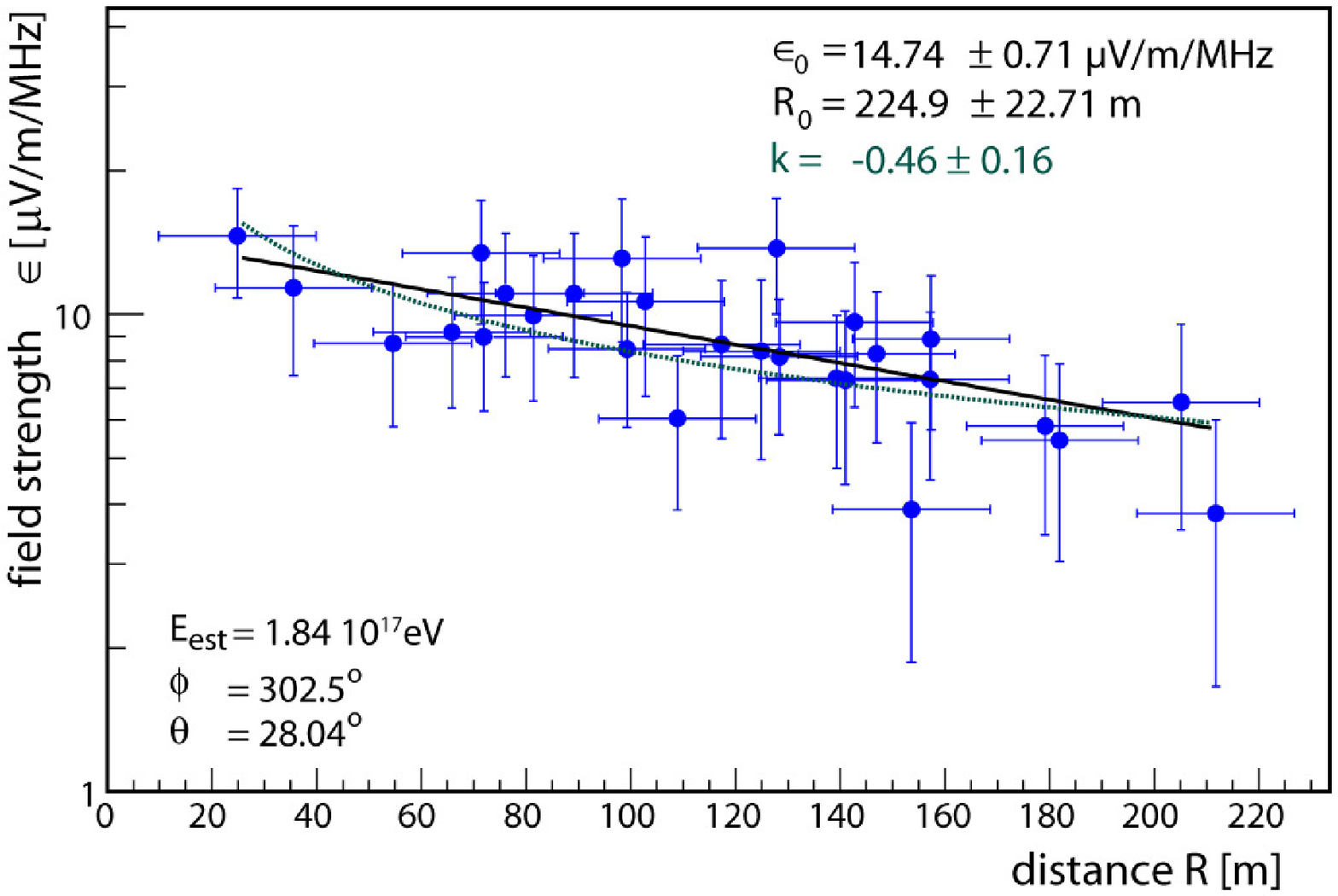}
  \includegraphics[width=0.48\textwidth]{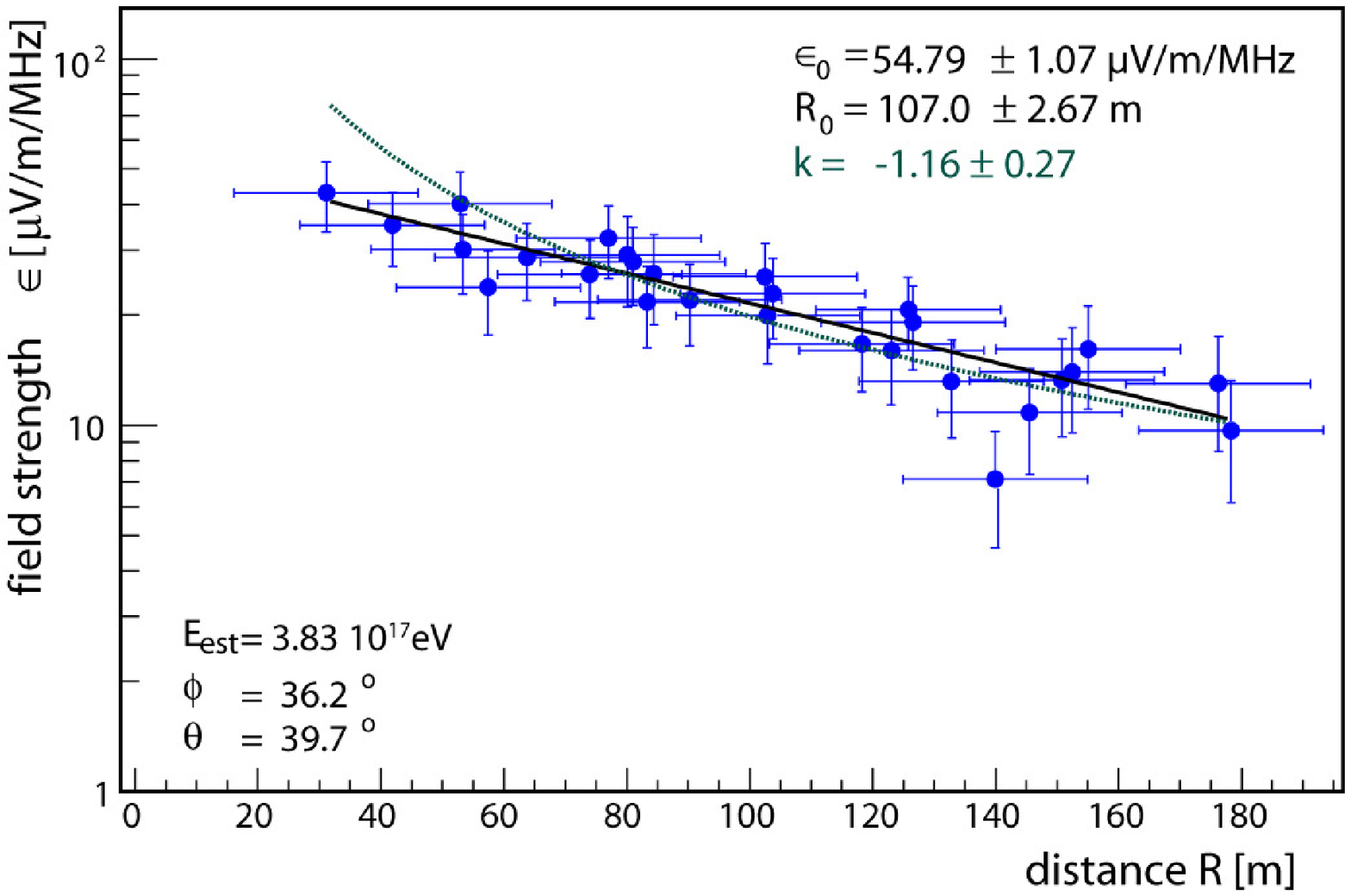}
\end{center}
\caption{Lateral distributions reconstructed from single antenna signals, 
shown for two individual showers. The full and dashed lines show the result of an exponential and power law fit, respectively. The left panel shows the same event as used for figure~\ref{ccbeam15-fit} and the right panel as used for figure~\ref{upsample}.}
\label{FigLatex}
\end{figure}

To investigate the lateral behavior of the radio signal in individual events
an exponential function 
$\epsilon=\epsilon_0\cdot\exp\left( -R/R_0 \right)$ is used to describe 
the measured field strengths. The fit contains two free parameters, where the scale 
parameter $R_0$ describes the lateral slope and $\epsilon_0$ the extrapolated 
field strength at the shower axis at observation level.
The number of antennas used for the determination of the two fit parameters 
can change from event to event, because an antenna can be flagged 
if the signal is disturbed by too high noise or by a technical malfunction 
at the time of the event. 
For comparisons, the signals of each event is also described by the best fitting power law $\epsilon=\epsilon_p\cdot R^k$.

Examples of events including the resulting lateral 
field strength functions are shown in figure~\ref{FigLatex}. 
The error bars of the individual values in x- and y-direction are derived  
from the field strength and distance estimation, respectively.
The uncertainty of the field strengths is calculated in a conservative way and includes the 
uncertainty of the antenna calibration and a value appointed to the maximum contribution 
of the noise. 
To avoid implicit assumptions on the lateral behavior of the field strength, we consider each 
antenna data as an individual measurement. But by that, the errors 
in x-direction and the uncertainties from the antenna calibration are not 
independent from other data points in an individual event and the significance of the fit results
will be slightly reduced.  
The displayed showers are typical for almost all investigated events of which
the distribution could be apparently described by an exponential function. 
The power law function in most cases tends to overestimate the lateral behavior close to the shower core, which also lead to a slightly worse quality of the fits ($\chi^2/ndf$). Therefore, in the further analysis mainly the results of fits with the exponential function is considered.

\subsection{Unusual Lateral Profiles}

\begin{figure}[ht]
 \centering
  \includegraphics[width=0.48\textwidth]{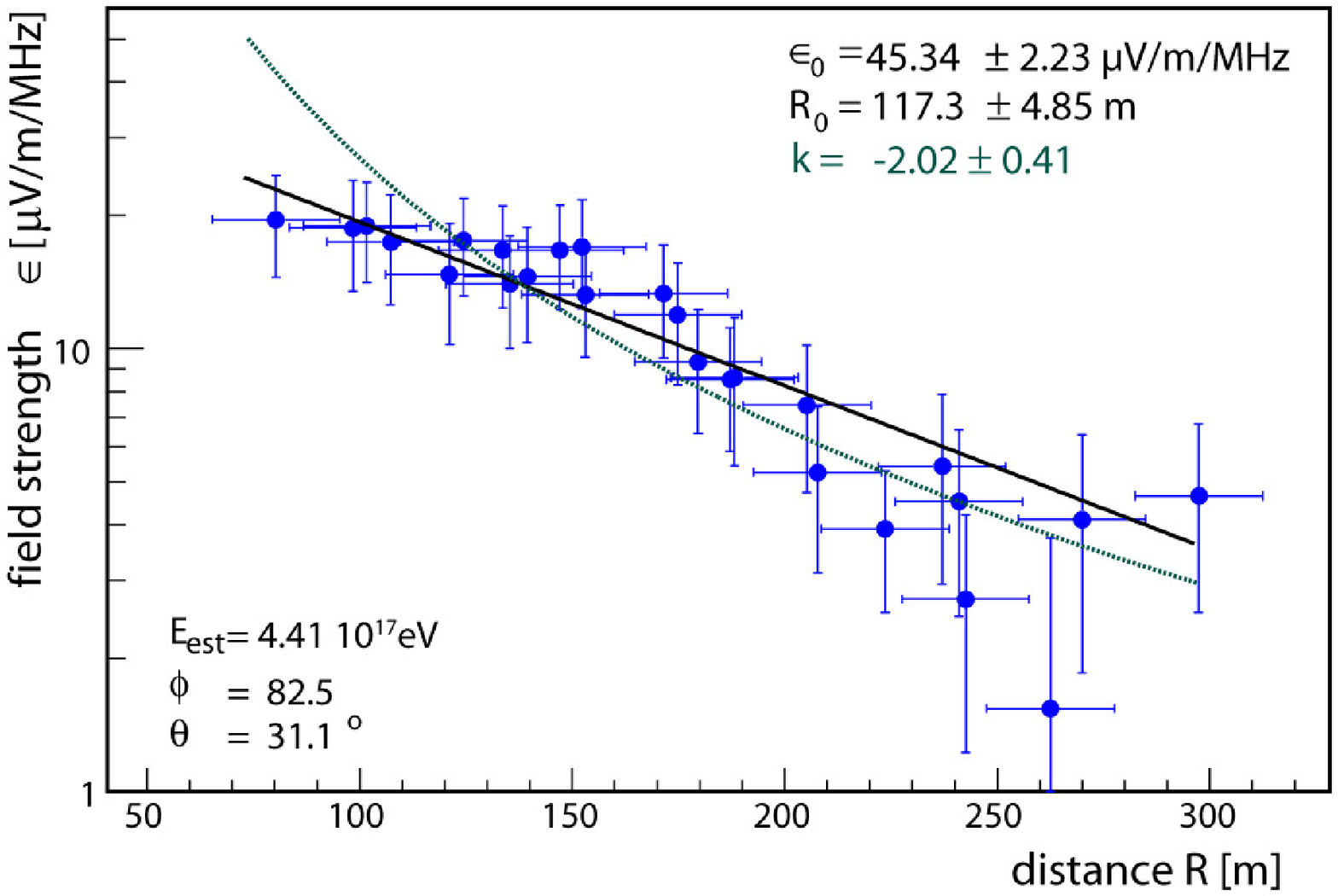}
  \includegraphics[width=0.48\textwidth]{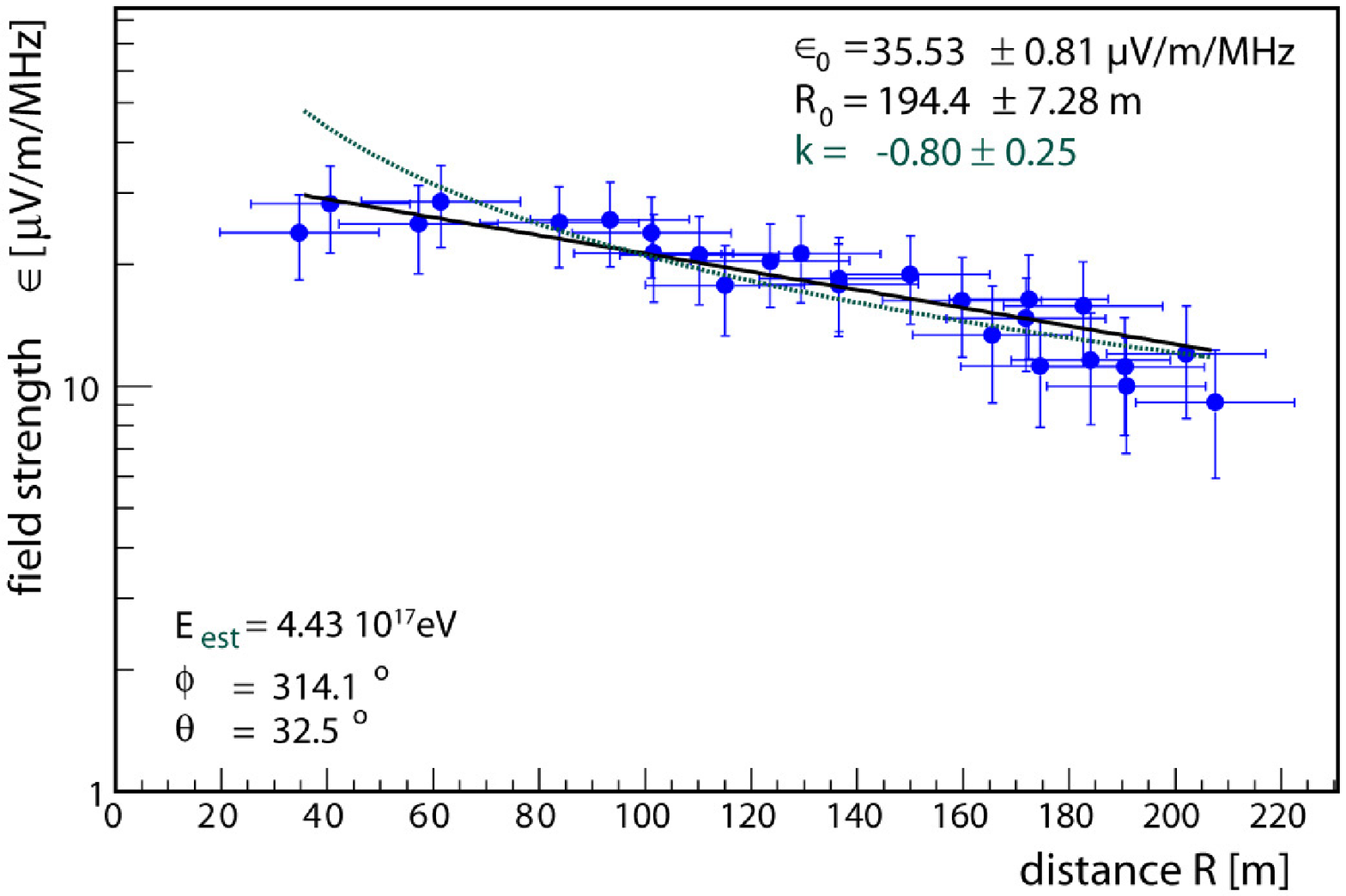}
  \includegraphics[width=0.48\textwidth]{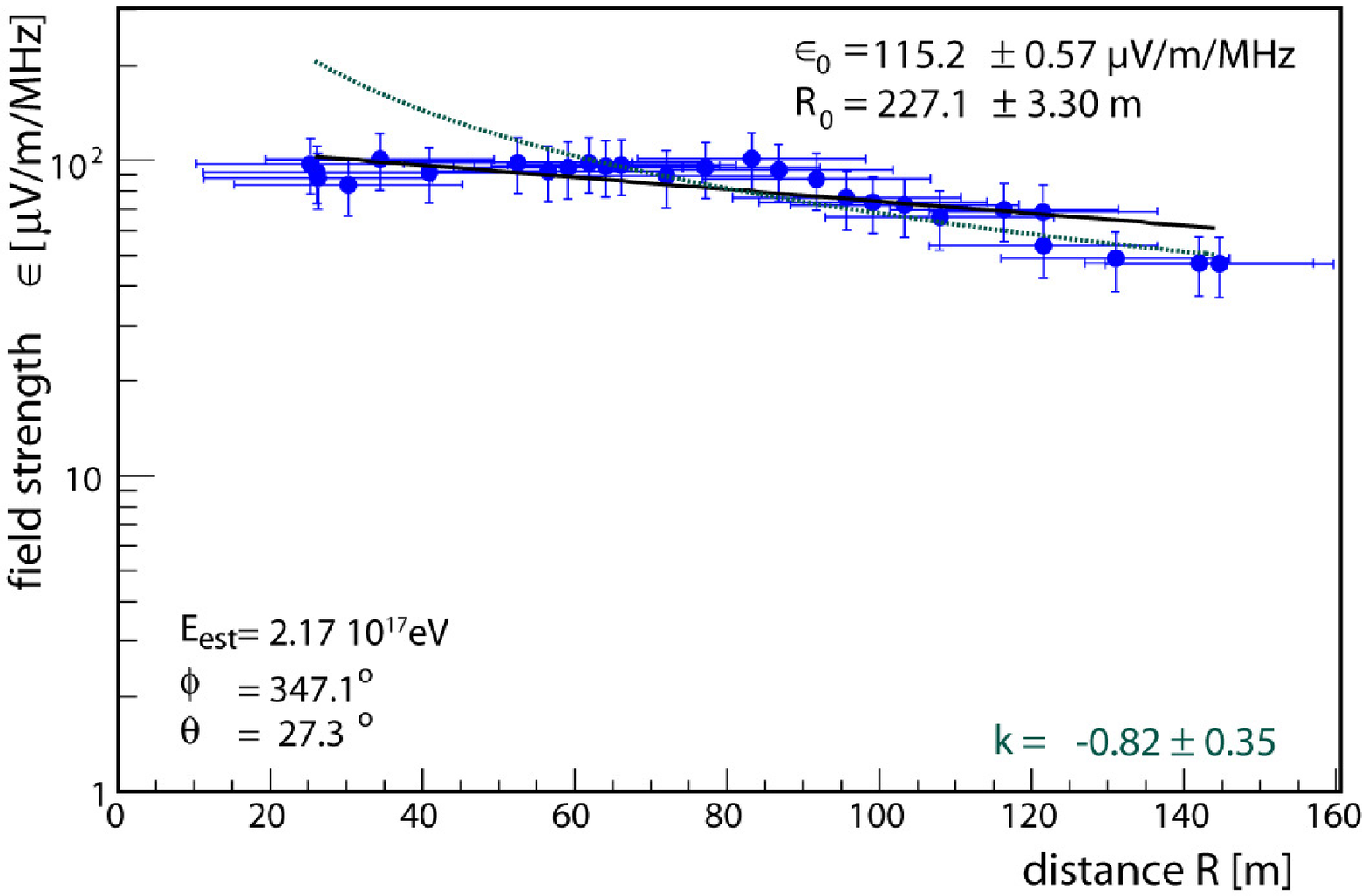}
  \includegraphics[width=0.48\textwidth]{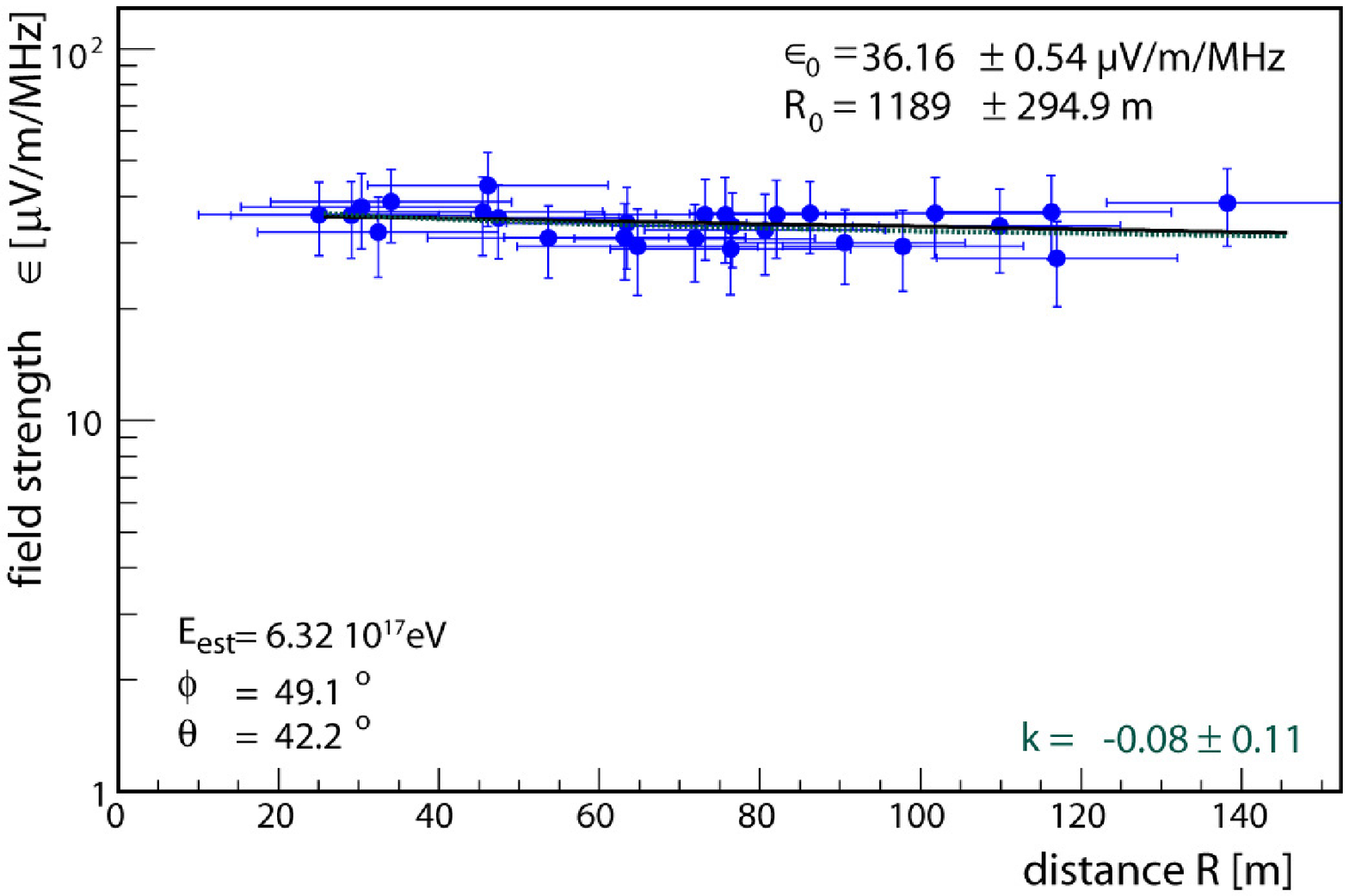}
   \caption{\label{abnormal} Same as 	Fig.~\ref{FigLatex}, but for four lateral distributions with unusual shapes. Discussion see text.
    }
\end{figure}

For roughly 20\% of the events lateral distributions have been found
which do not show a clear exponential or power law fall-off. 
Four of these special distributions are shown in figure~\ref{abnormal}.  

The shower displayed top left shows apparently a behavior as
others do, but there appears a flattening for small distances.  
Maybe, it could be considered that a fit of two functions 
with a break at $\approx 140\,$m would better describe the lateral profile. 
There are about 15 events that exhibit such a slope change
to a flatter lateral distribution close to the shower axis. The shower 
displayed top right shows another example of such events, but here the 
flattening occurs at $\approx 100\,$m. 

Also, for the shower in the bottom left panel
a slope change at a distance of about $90\,$m
can be seen. In addition, this particular shower shows a lateral
distribution that is very flat for the distance range up to $90\,$m.

The even more extreme case of a flat lateral distribution is shown in the  
bottom right panel of figure~\ref{abnormal}. Over the whole distance range that could
be measured, there is almost no fall-off in field strength. 

It should be remarked that at field strengths above $5\,\mu$V/m/MHz 
the ambient noise background cannot affect the measurement and for none of the investigated events a saturation of an antenna occurred.  
Also no known instrumental or selection effects can explain such shapes, and no strange 
environmental conditions like a thunderstorm appeared during these events.
However, for a statistically reliable analysis, too few of such flat lateral profiles have been measured so far.

\subsection{The Scale Parameter R$_0$}

\begin{figure}[t]
  \centering
  \includegraphics[width=0.98\textwidth]{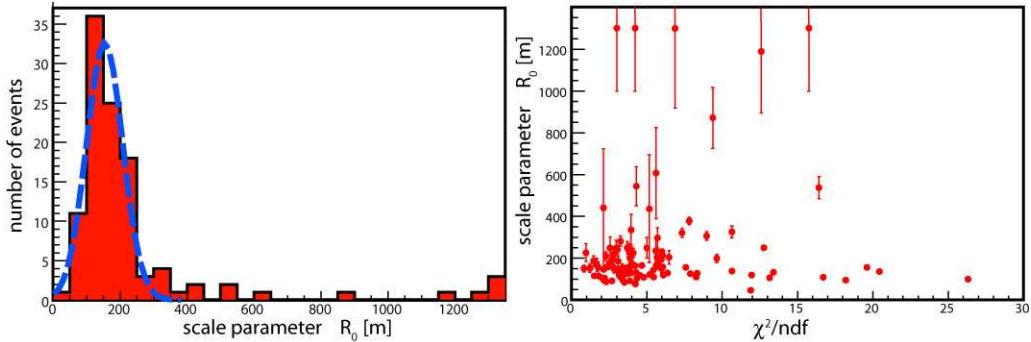}
  \caption{\label{fig:histo:scale}
    Distribution of the scale parameter $R_0$. There are four events set to
    $R_0=1300\,$m whose actual $R_0$ values are higher, 
    i.e.~these are very flat events. 
    The dashed line displays a Gaussian fit to the distribution in the 
    range $R_0=0-300\,$m resulting in $\bar{R_0} = 157\,$m with a width of $54\,$m. 
    The right panel shows the goodness-of-fit parameter in correlation with the obtained $R_0$.}
\end{figure}

The scale parameter R$_0$ describes the slope of the lateral distribution.
Most of the showers have a scale parameter smaller than $250\,$m (figure~\ref{fig:histo:scale}).
There are some showers with extremely large scale parameter, $R_0>1300\,$m, that
are set in the plots to $R_0=1300\,$m.  The distribution peaks at a scale
parameter of $R_0\approx 125\,$m, but due to the $\approx 10$\% of very flat 
showers the median value is $R_0=155\,$m for the complete sample.
Fitting a Gaussian function to the distribution of the scale parameter in the range of 
$0-300\,$m, i.e. neglecting the flat events, a mean value of $\bar{R_0} = 157\,$m 
with a width of $54\,$m is obtained (figure~\ref{fig:histo:scale}).

As the uncertainty in signal strength is estimated in a conservative way and by that the errors 
of individual measures are not independent, the fit quality in terms of $\chi^2/ndf$ is reduced. 
The right panel of figure~\ref{fig:histo:scale} displays the individual $\chi^2/ndf$ in relation 
to the obtained scale parameter $R_0$, where no correlation is found, not even for the unusual 
flat profiles.   

\begin{figure}[ht]
  \centering
  \includegraphics[width=0.98\textwidth]{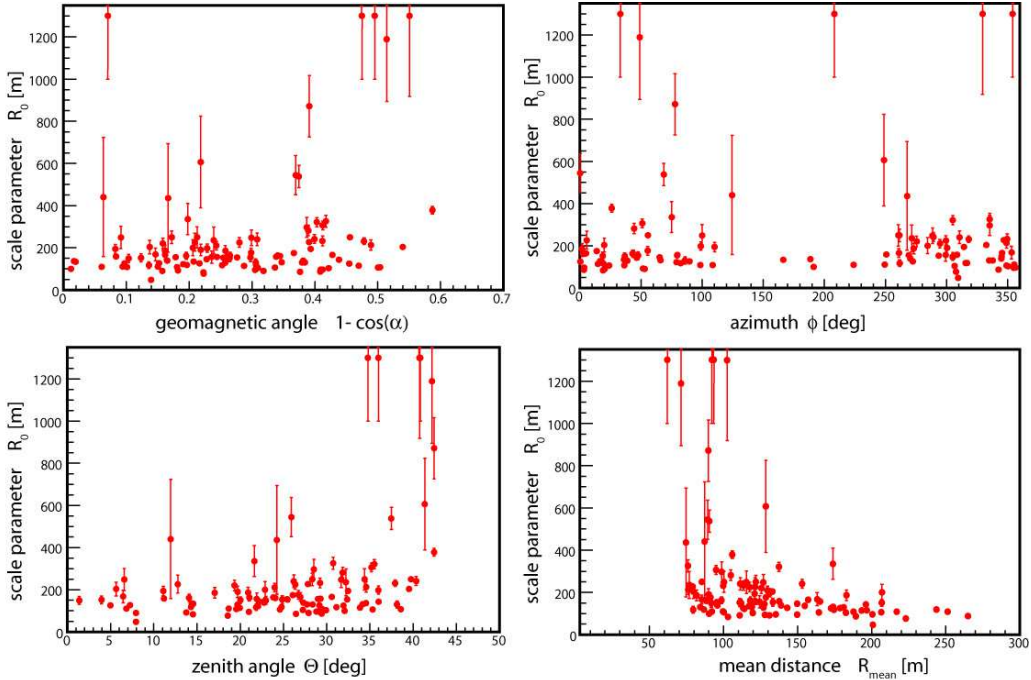}
  \caption{\label{fig:corr:R0} Relations 
    of the scale parameter $R_0$ with geomagnetic, azimuth, and zenith
    angle of the showers as well as with the mean distance of the antennas to the 
    shower axes. The error bars show the uncertainty obtained by the fit 
    with the exponential function.
    }
\end{figure}

For a more detailed study of the lateral distributions the properties of the scale 
parameter and possible correlations with EAS parameters have been investigated. 
In case of the LOPES experiment this can be done easily, as the shower parameters are
obtained from the KASCADE-Grande measurements.
For example, the shower arrival direction may play a role for the obtained shape
of the lateral distribution. In order to test this, the scale parameter
is correlated with the geomagnetic angle $\alpha$ (using an
($1-\cos{\alpha}$)-functional dependence\footnote{Using the cc-beam it was found 
that this functional form describes best the correlation 
of the field strength measured with LOPES with the geomagnetic field; at least for the east-west 
polarization component~\cite{Horn07}.}), the azimuth angle $\phi$, and the zenith angle
$\theta$ (figure~\ref{fig:corr:R0}). 
A correlation is found neither between the geomagnetic angle and $R_0$ nor 
between the azimuth angle and $R_0$. 

The situation is different when the scale parameter is correlated with
the zenith angle of the incoming primary cosmic ray (lower left panel 
of figure~\ref{fig:corr:R0}). Here a tendency towards larger values of the 
scale parameter is seen for inclined events. Expectations from simulations~\cite{Huege05}
that very inclined showers in general exhibit a larger scale parameter cannot 
be directly proven, as the cut for the zenith angle larger
than $44^\circ$ inhibits any further conclusions at the moment.
\begin{figure}[ht]
  \centering
  \includegraphics[width=0.55\textwidth]{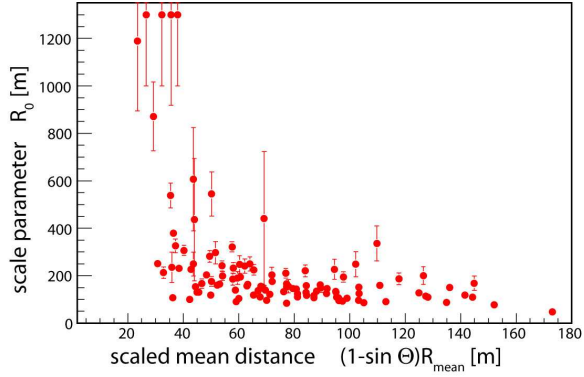}
  \caption{\label{fig:corr:Rsc} Relation 
    of the scale parameter $R_0$ with the mean distance of the antennas to the shower 
    axis scaled with the zenith angle of the axis.  
    }
\end{figure}

A clearer feature is seen, when the scale parameter is analyzed with respect to the 
corresponding mean distance to the shower axis of all antennas participating in an 
individual event (figure~\ref{fig:corr:R0}, bottom right panel). 
It is obvious that for a fraction of the events the lateral profile gets flatter when we measure closer to the 
shower center. In particular, all the very flat events have a mean distance 
below $R_0\approx 80\,$m.
This is connected with the fact that we see showers which flatten towards the shower 
core, as discussed in the previous section. 

An even more pronounced dependence of the scale parameter of flat events is seen when 
the mean distance is combined with the zenith angle information in the form 
$R_{\rm mean}^\prime = (1-\sin{\Theta})\cdot R_{\rm mean}$. Figure~\ref{fig:corr:Rsc} 
shows clearly that the probability of a flattening increases when the shower is inclined 
and when measured closer to the shower axis. Cutting events with 
$R_{\rm mean}^\prime < 50\,$m and fitting again a Gaussian function to the distribution 
of the scale parameter in the range of $0-300\,$m, the mean value and width decrease 
to $\bar{R_0} = 149\,$m and $\sigma=50\,$m, respectively. 
But as not all events with small $R_{\rm mean}^\prime$ show a flattening the reason 
is still unclear and further investigations with larger statistics are required. 

In addition, a comparison of the scale parameter with the estimated 
primary energy (figure~\ref{fig:comp:E0}, left panel) has been performed, 
where no correlation could be observed. 
This is also the case (not shown here) when the correlation 
of $R_0$ with the lateral scale parameter\footnote{The so-called `age' of the 
extensive air shower, which is reconstructed by using the particle density measurements 
of KASCADE-Grande.} of the particle component of the EAS is investigated.

\subsection{The Absolute Field Strength $\epsilon_0$}

The second parameter from the fitting procedure $\epsilon_0$ 
represents the field strength at the shower axis. 
It is obtained by extrapolating the measured lateral profile to the shower axis.  
\begin{figure}
  \centering
  \includegraphics[width=0.98\textwidth]{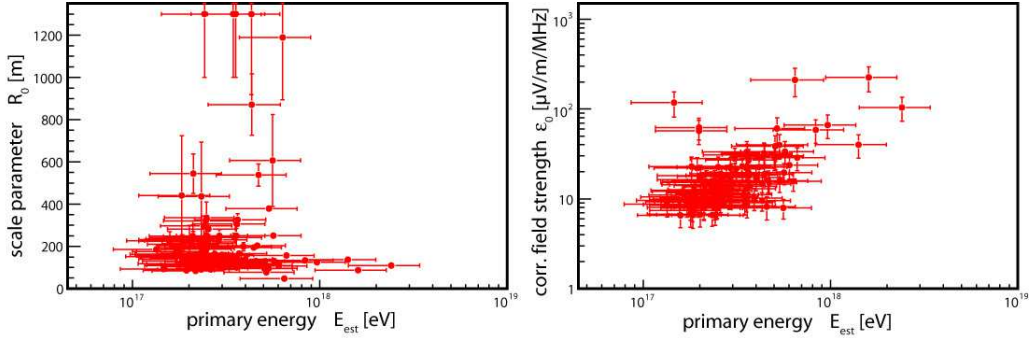}
  \caption{\label{fig:comp:E0}
    The scale parameter $R_0$ (left panel) and the field strength at the shower 
    center $\epsilon_0$ (right panel) as a function
    of the estimated primary energy $E_{\rm est}$. $\epsilon_0$ is the
    result of a fit of the lateral distribution of single events corrected for 
    the geomagnetic angle.}
\end{figure}

For the investigated events the fit of the field strength
$\epsilon_0$ yields values as expected from earlier 
investigations~\cite{Horn07}. Figure~\ref{fig:comp:E0}, right panel, shows 
the relation of $\epsilon_0$ versus the estimated primary energy $E_{\rm est}$,
where $\epsilon_0$ is corrected for the geomagnetic angle with a 
(1-$\cos{\alpha}$)-dependence.
The uncertainty of the estimated primary energy, as well as the systematic 
uncertainty in the amplitude calibration is taken into account. 
A clear dependence of $\epsilon_0$ with the primary energy is visible 
following a power law $\epsilon_0\propto E^\kappa$ with the index $\kappa$ close 
to $1$. This is expected if there is a coherent emission of the radio signal 
during the shower development. The absolute scale of the radio field strength 
as well as the extent and source of fluctuations are still under investigation
and will be subject of a forthcoming paper.

\subsection{Discussion}

The lateral distributions of the electric field strength measured in 
individual events are described by exponential functions resulting in two parameters per EAS which are subject of further investigations.
For most of the events the lateral distribution of the measured field strengths can also be described by a power law, though these fits tend to overestimate the signal strengths closer to the shower axis. 
The choice of the exponential dependence is also driven by simulation studies 
which predict an exponential fall-off of the lateral distribution~\cite{Huege07}.
Results from LOPES data analysis investigating the dependence of 
the cc-beam values with mean distance of the antennas to the shower 
axis~\cite{Horn07} have also shown an exponential behavior, as well as 
early investigations leading to the parametrization of Allan 1971~\cite{Allan71}, 
and results from the CODALEMA experiment~\cite{Ardouin06}.

The scale parameter $R_0$ obtained from the fits has a relatively narrow peak
at $R_0\approx125$~m and a long tail with partially very large scale parameters. 
No direct dependence of $R_0$ on the geomagnetic angle, the primary energy and 
the azimuth angle could be found, which indicates that the shape of the lateral 
distribution is an intrinsic property of the radio emission in extensive air showers.
Inclusion of the tail of the distribution in figure~\ref{fig:histo:scale} leads to a 
mean value that agrees with the cc-beam based scale parameter in the parameterization 
results of LOPES~\cite{Horn07}, whereas the exclusion of the tail yields a scale
parameter that agrees with the parameterization of earlier data, e.g.~$R_0 \approx 110\,$m 
by Allan~\cite{Allan71}. One should have in mind that these earlier measurements compiled 
by Allan in ref.~\cite{Allan71} were performed in narrow bandwidths, and the 
dependence of the lateral distribution on frequency is still an open issue. 
The CODALEMA experiment has measured a few individual events with scale parameters 
in the order of around $200\,$m~\cite{Ardouin06} for the absolute signal measured in both 
polarization directions.

Roughly 10\% of the investigated showers show very flat lateral distributions 
with scale parameters $R_0>300\,$m (examples shown in figure~\ref{abnormal}). 
It was found that these showers are preferably arriving with a larger inclination 
and have their core closer to the LOPES antennas, i.e.~the mean distance of the 
antennas to the shower axis is relatively small.
In approximately another $10$\% of the measured events we recognize a lateral 
distribution with a flattening towards the shower center, where the fit to the 
distribution reveals only a slight increase of the scale parameter $R_0$.
Combining these two observations, the very flat showers are probably events where we 
only measure the flat part close to the axis. 
The observation of a slight increase of the average $R_0$ with zenith angle can serve 
as a confirmation that for inclined showers the enlarged footprint on ground might 
enhance the effect.
It is not expected that an exponential or a power law will fit the lateral distribution of radio signals perfectly, because both lead to a discontinuity at the shower axis. 
Therefore a modification for small distances from the shower axis is to be expected. 
However, it has to be mentioned that the flattening is observed only in a small 
fraction of the measured events and these events are not distinguishable from the others.
For the majority, a clear exponential behavior 
of the lateral distribution of the radio signal is observed. 
Instrumental, background or environmental (man-made or weather) effects can be excluded as the cause for the flattening of the radio profiles.
Hence, these experimental clues are remarkable and require
further investigations with higher statistics. 
Interesting is the fact, that also the CODALEMA experiment reported about a few events with such a flat lateral behavior close to the shower center~\cite{Laut08}.

A thorough understanding of the emission process is needed to understand the experimental findings. Besides the exponential lateral behavior predicted by the geo-synchrotron model~\cite{Huege08}, there are results from simulations which describe the lateral slope by a power law~\cite{Scholten08} and earlier calculations indicating a change of slope toward the shower axis~\cite{Huege03}. No theoretical approach investigated so far, however, predicts slope parameters in the order of $500-1000\,$m in any range of lateral distance.

Case studies of special lateral distributions are presently limited
in statistics and demand a much larger data set. 
However, the observed flattening of the lateral distribution of the radio signal to the shower 
center is an unexpected feature, in particular as it is only observed for a fraction of the events. Detailed information about the 
corresponding air shower is required to understand the measured lateral profiles. 
This corroborates the unique quality of 
the coincide measurements of LOPES with KASCADE-Grande.
Also the role of possible polarization effects and the effect of the chosen observing frequency 
window on the radio lateral distribution have to be considered. 

The correlation of the estimated field strength at the shower axis (fig.~\ref{fig:comp:E0}) 
with the primary energy is a measure of the expected coherence of the radio emission in EAS. 
Due to the high detection threshold for the radio emission from EAS
and the limit in the energy range for the KASCADE-Grande experiment
most of the showers in this work are in a very narrow range of the primary energy. 
In LOPES radio data parameterizations~\cite{Horn07} the power law index resulted 
to $\kappa=0.95\pm0.04$. 
Also from simulations a relation is expected that can be described with an index 
$\kappa$ for a range between 1 and 0.75~\cite{Huege05} changing with the depth 
of the particle number maximum $X_{\rm max}$ as a function of primary energy.
Hence, the obtained correlation
supports the expectation of a scaling of the field strength
$\epsilon_0$ with primary energy and therefore a coherent emission.

The measured scale parameter of minimum $R_0 = 100\,$m at an energy of around 
$10^{17}\,$eV can be used to perform a rough estimate of the needed grid size for 
a radio antenna array measuring EAS at higher energies, i.e.~planned in the frame of the Pierre Auger Observatory~\cite{vdBerg09}.
Assuming the exponential behavior with $R_0=100\,$m and the linear energy 
dependence of the signal, the distance between two antennas may be increased to
up to $230\,$m to measure an equal field strength when increasing the primary energy 
by a factor of ten.

\section{Conclusions and Outlook}

The described studies on lateral distributions of the radio signal in air showers 
are based on coincidence measurements of the LOPES radio antenna array and the air 
shower experiment KASCADE-Grande. This is a unique combination of two different 
detection techniques for EAS investigations.  
Furthermore, the obtained data of both experiments are well calibrated and can 
therefore be used to compare the properties of the radio emission with EAS 
parameters in detail. Such investigations aim to fully understand the radio emission 
in extensive air showers for primary energies below $10^{18}\,$eV.
In the frame of this work a calibrated data set of showers has been analyzed to study the 
lateral distribution of the radio emission in individual air showers.

The applied analysis method uses up-sampled signals from individual antennas to
derive the electric field strength per unit bandwidth $\epsilon$ (40-80$\,$MHz and east-west polarization cmponent, only). With the help of the
reconstructed shower parameters from KASCADE-Grande the lateral
distribution of the radio signal in EAS for individual events is obtained. 
Here, the uncertainty of the amplitude calibration and 
an antenna-by-antenna event-by-event background calculation enters into 
the uncertainty of the determined field strength $\epsilon$. The 
uncertainty of the reconstructed geometry of the shower contributes to 
the uncertainty for the lateral distance $R$ of each antenna. 
For 110 showers which have a large enough radio signal in all antennas the lateral 
shape of the signal as well as correlations with global shower parameters, like
direction and energy, were studied.

Exponential as well as power law functions were applied to the measured profiles, where the first one slightly better describes the data. 
The exponential function has two free parameters, the field strength 
$\epsilon_0$ at the shower axis, and the scale parameter $R_0$.
The scale parameter $R_0$ is a quantity that describes the slope of the lateral
profile. The scale parameter distribution shows a peak
value of $R_0\approx125$~m and has a tail with very flat lateral distributions. 
Excluding the flat events a mean value of $\bar{R_0} \approx 150\,$m with a width 
of $\sigma=50\,$m was obtained. No direct evidence for a dependence on the shower 
parameters azimuth angle, geomagnetic angle, and primary energy could be found.  
This indicates that the lateral profile is an intrinsic property of the radio emission 
and the shower development. Comparing the obtained scale parameter with published 
values of earlier experiments, a good agreement has been found.
Studying the lateral distributions in individual events, approximately 20\% of the
studied showers show a very flat lateral distribution or exhibit a flattening 
towards the shower center. Preferably, such showers arrive under larger zenith angle and 
axes are close to the antennas. But, there are too few such showers measured to derive any
significant correlation with specific shower parameters.  
However, the result is an indication that there might be 
(at least for a part of the showers) additional, not yet understood effects of 
the radio emission during the shower development.
In the near future, besides higher statistics delivered by LOPES, the LOFAR project 
with its hundreds of antennas will provide detailed information (including polarization effects) 
on the lateral distribution of the radio signal in EAS. 

The present analysis concentrates on experimental findings, only, but for the understanding of the radio emission observed in EAS a comparison with 
detailed Monte-Carlo simulations on an event-to-event basis will be very important. 
This will be possible soon because of the performed amplitude calibration of 
LOPES and the estimate of the field strength at individual antennas. 
Due to a new simulation strategy, using more realistic air shower models 
derived from per-shower CORSIKA~\cite{cors} simulations including radio 
emission simulations with REAS2~\cite{Huege07} and a simulation of the antenna response, detailed comparisons between 
LOPES measured events and simulations can be performed.   

The future goal of the radio detection technique lies in a large scale application 
with hundreds of antennas. 
The quadratic dependence of the received radio power on primary energy 
and the comparatively (with respect to the electromagnetic component of EAS) 
large lateral scale parameter will make radio detection a cost effective method for 
measuring the longitudinal development of air showers of the highest energy cosmic rays. 

\begin{ack}
LOPES has been supported by the German Federal Ministry of Education
and Research. The KASCADE-Grande experiment is supported
by the German Federal Ministry of Education and
Research, the MIUR and INAF of Italy, the Polish Ministry
of Science and Higher Education and the Romanian Ministry
of Education and Research.
\end{ack}

\end{document}